\documentclass{article}

\usepackage{mypreamble} 
\usepackage{csquotes}
\usepackage[style=numeric,backend=bibtex]{biblatex}
\usepackage{biblatex}[=3.2]

\usepackage{dsfont}
\usepackage{scalerel}
\usepackage{marginnote}
\usepackage{mwe}
\usepackage{wrapfig}
\usepackage{tikz}
\usetikzlibrary{automata, positioning}
\usetikzlibrary{shapes.geometric}
\usetikzlibrary{calc}
\usepackage[normalem]{ulem}
\usepackage{pgfplots}
\pgfplotsset{compat=1.17}
\definecolor{MyShadeOfBlue}{rgb}{0.0, 0.0, 0.6}

\title{Sparse canonical correlation analysis for multiple measurements with latent trajectories}
\author{Nuria Senar\footnote{Corresponding author. E-mail: n.senarvilladeamigo@amsterdamumc.nl. \\ \indent \indent \indent \indent \indent \indent \indent \indent \indent \indent \indent \indent Postal address: Meibergdreef 9, Location J1B, 1105 AZ, Amsterdam, The Netherlands
}}
\author{Aeilko H.\ Zwinderman}
\author{Michel H.\ Hof}
\affil{\small Department of Epidemiology \& Data Science, Amsterdam School of Public Health, Amsterdam UMC, Amsterdam, The Netherlands}
\date{January, 2025}

\begin{document}
\maketitle

\begin{abstract}
    \noindent Canonical Correlation Analysis (CCA) is a widely used multivariate method in omics research for integrating high-dimensional datasets. CCA identifies hidden links by deriving linear projections of observed features that maximally correlate datasets. An important requirement of standard CCA is that observations are independent of each other. As a result, it cannot properly deal with repeated measurements. 
    Current CCA extensions dealing with these challenges either perform CCA on summarised data  
    or estimate correlations for each measurement. While these techniques factor in the correlation between measurements, they are sub-optimal for high-dimensional analysis and exploiting this data's longitudinal qualities.

    We propose a novel extension of sparse CCA that incorporates time dynamics at the latent variable level through longitudinal models.
    This approach addresses the correlation of repeated measurements while drawing latent paths, 
    focusing on dynamics in the correlation structures.
    To aid interpretability and computational efficiency, we implement an $\ell_0$ penalty to enforce fixed sparsity levels.

    We estimate these trajectories fitting longitudinal models to the low-dimensional latent variables, 
     leveraging the clustered structure of high-dimensional datasets, thus  exploring shared longitudinal latent mechanisms.
    Furthermore, modelling time in the latent space significantly reduces computational burden.
    We validate our model’s performance using simulated data and show its real-world applicability with data from the Human Microbiome Project. 

    Our CCA method for repeated measurements enables efficient estimation of canonical correlations across measurements for clustered data. Compared to existing methods, ours substantially reduces computational time in high-dimensional analyses as well as provides 
    longitudinal trajectories that yield interpretable and insightful results.

\end{abstract}


\section{Introduction}\label{sec:intro}

Biomedical data, such as genetic or methylation data, is often high-dimensional, with the number of features far exceeding the number of samples. Analyzing this type of data requires methods capable of addressing challenges posed by high dimensionality, including collinearity and interpretability. Advances in technology have improved the ability to collect high-throughput biomedical data through enhanced measurement instruments and reduced patient burden, enabling the collection of multiple measurements from the same individual. These multiple (or repeated) measurements can provide valuable insights into the time dynamics of the underlying mechanisms, revealing longitudinal associations that might otherwise remain hidden. These associations offer a deeper understanding of biological or clinical processes. However, repeated measurements from the same patient often introduce dependencies across observations. Furthermore, the data may include missing measurements for some individuals or be recorded at uneven intervals, resulting in sparsely and irregularly observed data.  

Integrative methods, such as Canonical Correlation Analysis (CCA), combine multiple datasets through maximally correlated linear projections, typically refered to as latent vectors. By uncovering shared variation between datasets, CCA provides insights into the underlying relations that exist between the variables from these data sources. Sparse extensions of CCA further enhance this exploration by identifying the most relevant variables, helping to mitigate collinearity and focus on key contributors to the observed associations.

To deal with repeated measurements in CCA, Hao et al. (2017) and Du et al. (2019) \parencite{lccTempCons_hao_2017, multiTaskSCCA_du_2019} proposed to  estimate a separate sparse CCA for each unique time point, using \textit{fused}-type penalties to ensure consistency in variable selection across time points. Although these approaches are suitable for high-dimensional data, they require correspondence and regularity between measurements across datasets. Alternatively, Lee et al. (2023) \parencite{lee_longitudinal_2023} proposed performing CCA on summarised data for each measurement, effectively addressing irregularly and sparsely observed data, yet remaining unsuitable for high-dimensional data. The above methods do not focus on the dynamics of latent mechanisms and allow the canonical weights to vary across measurements, which may introduce challenges, such as feedback between the canonical weights and latent variables may lead to non-identifiability of their separate contributions. This, in turn, leaves the shared underlying dynamics of correlated feature groups unexplored. Overall, these methods either fail to leverage the longitudinal structure of the data, cannot handle high-dimensionality, or are limited by irregularly observed measurements.

We address all three points above by fixing the canonical weights across all measurements and modelling the longitudinal dimension in the latent space, thus tracking the evolution of these correlated underlying mechanisms. Since the latent space is low-dimensional, this approach maintains low computational times. To the best of our knowledge, no existing methods focus on describing latent trajectories connecting high-dimensional datasets.

We validated our model using simulations, where it successfully identified the simulated latent trajectories and the main contributors. Additionally, we applied our method to real data from the Human Microbiome Project (HMP) \parencite{hmp_zhou_2019}, uncovering latent paths that link gene expression and operational taxonomic units (OTUs) from gut microbiome data for insulin-sensitive (IS) and insulin-resistant (IR) patients. Notably, our algorithm described latent paths consistent with findings reported in existing literature.

\section{Methods}

Suppose we have data matrices $\mathcal{X}=(\boldsymbol X_1,\boldsymbol X_2,\ldots \boldsymbol X_n)^\intercal$ and $\mathcal{Y}=(\boldsymbol Y_1,\boldsymbol Y_2,\ldots \boldsymbol Y_n)^\intercal$ containing the repeated measurements from $n$ individuals stacked in long format. Let $\boldsymbol{X}_{i}\in\mathbb{R}^{m_i\times p}$ be the observations for individual $i$ consisting of $m_{i}$ measurements of $p$ features. Similarly, $\boldsymbol{Y}_{i}\in\mathbb{R}^{s_i\times q}$ is the observation matrix of $s_i$ measurements of $q$ features for $i$. The times at which the measurements from individual $i$ are obtained are given by $\bt_{x,i}\in\mathbb{R}^{m_{i}}$  for $\bX_i$ and $\bt_{y,i}\in\mathbb{R}^{s_{i}}$ for $\bY_i$. We denote $\bt$ to the vector of all measurements if both $\mathcal{X}$ and $\mathcal{Y}$ such that $\mathbf{t} = \bigcup_{i} \left( \mathbf{t}_{x,i} \cup \mathbf{t}_{y,i} \right)$. 

Importantly, the number of measurements from individual $i$ and their corresponding collection dates can differ between both data matrices. For instance, individual $i$ may have three measurements in $\mathcal{X}$ and five measurements in $\mathcal{Y}$, i.e.\ $m_i=3$ and $s_i=5$. Additionally, these measurements may have been collected at different moments, e.g.\ $\bt_{x,i} = (1, 3, 4)$ and $\bt_{y,i} = (1, 2, 3, 4, 6)$, which leaves $\bt_i = (1, 2, 3, 4, 6)$. Moreover, these vectors may be organised to represent different time frames or sequences, depending on the longitudinal path of interest. For instance, these can simply indicate measurement number (as in the above example), collection dates or dates relative to an event of interest, as shown  in section \ref{sec:hmp}.

\begin{figure}[!h]
\centering
    \begin{tikzpicture}
    \node[anchor=east] at (-0.2, 0) {$\mathcal{X}=$}; 

    \node[anchor=east] at (6.8, 0) {$\mathcal{Y}=$}; 

    \draw[fill=gray!20] (0, 2.2) rectangle (2, 1.7);
    \node at (1, 1.95) {\textbf{$\bX_1$}};
    \draw[fill=gray!20] (0, 1.5) rectangle (2, 0.4);
    \node at (1, 0.95) {\textbf{$\bX_2$}};

    \node at (1, 0) {\(\vdots\)};
    \node at (1, -0.8) {\(\vdots\)};
    \draw[fill=gray!20] (0, -1.5) rectangle (2, -1.8);    
    \node at (1, -1.65) {\textbf{$\bX_n$}};

    \draw[fill=gray!20] (7, 2.2) rectangle (9, 1.7);
    \node at (8, 1.95) {\textbf{$\bY_1$}};
    \draw[fill=gray!20] (7, 1.5) rectangle (9, 0.7);
    \node at (8, 1.1) {\textbf{$\bY_2$}};
    \node at (8, 0.2) {\(\vdots\)};
    \node at (8, -0.4) {\(\vdots\)};
    \draw[fill=gray!20] (7, -1) rectangle (9, -1.8);
    \node at (8, -1.45) {\textbf{$\bY_n$}};
    
    \draw[decorate, decoration={brace, amplitude=1.5}] (2.1, 2.2) -- (2.1, 1.6) node[midway, xshift=0.4cm] {\footnotesize$m_1$};
    \draw[decorate, decoration={brace, amplitude=1.5}] (9.1, 2.2) -- (9.1, 1.7) node[midway, xshift=0.4cm] {\footnotesize$s_1$};

    \draw[decorate, decoration={brace, amplitude=1.5}] (2.1, 1.5) -- (2.1, 0.4) node[midway, xshift=0.4cm] {\footnotesize$m_2$};
    \draw[decorate, decoration={brace, amplitude=1.5}] (9.1, 1.5) -- (9.1, 0.7) node[midway, xshift=0.4cm] {\footnotesize$s_2$};

    \draw[decorate, decoration={brace, amplitude=1.5}] (2.1, -1.5) -- (2.1, -1.8) node[midway, xshift=0.4cm] {\footnotesize$m_n$};
    \draw[decorate, decoration={brace, amplitude=1.5}] (9.1, -1) -- (9.1, -1.8) node[midway, xshift=0.4cm] {\footnotesize$s_n$};

    \draw[decorate, decoration={brace, amplitude=5pt}] (0, 2.3) -- (2, 2.3) node[midway, yshift=0.3cm] {\footnotesize $p$};
    \draw[decorate, decoration={brace, amplitude=5pt}] (7, 2.3) -- (9, 2.3) node[midway, yshift=0.3cm] {\footnotesize $q$};

    \end{tikzpicture}
    \caption{Data structure. \footnotesize The data is comprised of $n$ stacked data matrices of $p$ and $q$ features, respectively. The gray rectangles represent the multiple measures, $m_i$ and $s_i$ for each dataset, collected for each patient $i$. That is, for $\mathcal{X}$, the each rectangle is $\bX_i\in \mathbb{R}^{m_i\times p}$ and $\bY_i\in \mathbb{R}^{s_i\times q}$ for $\mathcal{Y}$.}
    \label{fig:data-structure}
\end{figure}
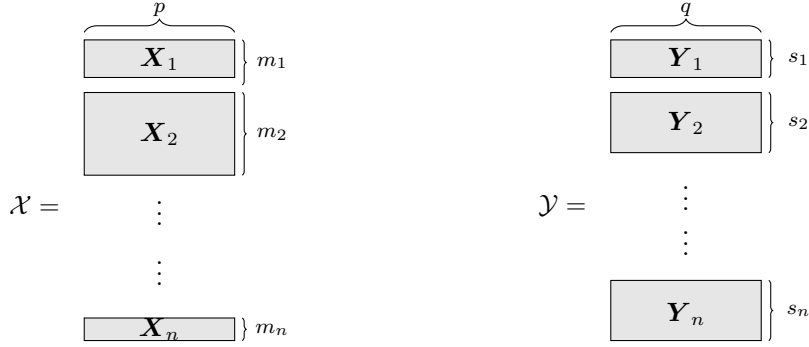

The main objective of CCA is to extract $K$ pairs of latent variables $\{( \boldsymbol{\eta}_1, \boldsymbol{\gamma}_1), (\boldsymbol{\eta}_2, \boldsymbol{\gamma}_2), \ldots, (\boldsymbol{\eta}_K,\boldsymbol{\gamma}_K)\}$ from both data matrices. Each latent variable is a linear combination of the features from a particular data matrix, that is, the $k^{th}$ pair of latent variables for individual $i$ is given by $\boldsymbol{\eta}_k = \mathcal{X} \mathbf{w}_{x,k}$ and $\boldsymbol{\gamma}_k = \mathcal{Y} \mathbf w_{y,k}$, where the weights $\mathbf{w}_{x,k}\in\mathbb{R}^{p}$ and $\mathbf{w}_{y,k}\in\mathbb{R}^{q}$ denote the contribution of each variable.

The weights $\bW_x=(\mathbf{w}_{x,1}, \ldots, \mathbf{w}_{x,K})$ and $\bW_y=(\mathbf{w}_{x,1}, \ldots, \mathbf{w}_{x,K})$ are chosen such that the canonical correlation between all pairs of latent variable is maximized, i.e.\ 
\begin{align} 
    (\widehat{\bW}_x, \widehat{\bW}_y) = &\arg\min_{ \bW_x,  \bW_y} \sum_{i=1}^n \bigl( {\mathbf{X}_i}  \bW_{x} -      {\mathbf{Y}_i}  \bW_{y}   \bigr)^2\label{eq:paper5_E1}
\end{align}
\noindent
under the restriction that the columns in the sets $(\bgamma_1, \ldots, \bgamma_K)$ and $(\boldsymbol{\eta}_1, \ldots, \boldsymbol{\eta}_K)$ are orthogonal.

In this paper, we focus on the Thresholded Ordered Sparse CCA (TOSCCA) \parencite{senar_toscca_2024} algorithm to obtain sparse weights for both datasets. Note, however, that our extension is general and can be applied to other CCA estimation procedures.

\subsection{TOSCCA}\label{sec:toscca}
 
TOSCCA expands on the standard CCA problem by re-framing the estimation to a least squares problem via the Nonlinear Iterative Partial Least Squares Algorithm (NIPALS) and using soft-thresholding which explicitly sets sparsity levels. 

The TOSCCA algorithm considers the objective of CCA as regression problem that can be solved by sequentially estimating pairs of latent variables. Each pair is estimated using the NIPALS algorithm \parencite{nipals_wold}. NIPALS starts by randomly initialising one of the canonical vectors, $\bW_x^{(0)}$. Then,  estimating $\bW_y$ given  $\bW_x^{(0)}$, reduces to a least squares problem (equation \eqref{eq:paper5_E1}). This process is then repeated until convergence of some tolerance measure. To deal with high-dimensionality, TOSCCA imposes sparsity via soft-thresholding using threshold parameters, $p_x$ and $p_y$, indicating the number of nonzero coefficients to estimate. 

TOSCCA's penalisation allows us direct control on the number of nonzero weights in both $\bW_x$ and $\bW_y$, improving interpretation and speeding up the computation as we do not have to search for the penalty value that corresponds to a particular number of nonzero weights.  

\subsection{Time dynamics}\label{sec:tim_dyn}

Since the number and timing of measurements for each individual can differ, we propose to add the longitudinal nature of the data to the CCA objective by assuming that the time dynamics only directly affect the shared hidden path. That is, the longitudinal trajectories affect the observations only through latent variables. Modelling the time dynamics in the latent space will not only help interpretation of the model, it also keeps the computation burden low, as the latent space is low-dimensional. We assume weights $\bW_x$ and $\bW_y$ to be fixed over time. Based on both assumptions, for each component $k = 1, \dots, K$, the observation from the $i^{th}$ individual from dataset $\mathcal{X}$ at time $t\in \mathbb{R}$ is a $p$-dimensional vector that follows 

\begin{equation}\label{eq:obs_lv_long_g}
\begin{split} 
    \bx_{i}(t) & = \eta_{i}(t)\bW_x^\intercal + \boldsymbol{\varepsilon}_{x,i}(t) \\
		& = f(t; \boldsymbol{{\theta}}_{\eta})\bW_x^\intercal + \boldsymbol{\varepsilon}_{x,i}(t) ,
\end{split}
\end{equation}

\noindent
where $\boldsymbol{\varepsilon}_{x,i}(t)$ is white noise.

Similarly, the observation at time $t \in\mathbb{R}$ for individual $i$ from dataset $\mathcal{Y}$ is a $q$-dimensional vector which can be described by

\begin{equation}\label{eq:obs_lv_long_e}
\begin{split} 
    \by_{i}(t) & = \gamma_{i}(t)\bW_y^\intercal + \boldsymbol{\varepsilon}_{y,i}(t) \\
		& = f(t; \boldsymbol{{\theta}}_{\gamma})\bW_y^\intercal + \boldsymbol{\varepsilon}_{y,i}(t) ,
\end{split}
\end{equation}
\noindent
where $\boldsymbol{\varepsilon}_{y,i}(t)$ is white noise.

The function $f(\cdot)$ describes the underlying mechanism linking both sets of observations through time. We assume this mechanism to follow the same model for both datasets. This is in line with the Probabilistic CCA \parencite{bach_probabilistic_2005}, where a single latent variable generates both datasets. Even though we separate this latent variable into two, $\boldsymbol{\eta}$ and $\bgamma$, the same principle applies.

For instance, $f(\cdot)$ may follow a mixed effect (LME) model with random intercepts and fixed effects for time for the latent trajectories, i.e.\ $f(t;\boldsymbol{\theta}_{\eta}) = {\alpha}_{x,i} +  {\beta}_{x} t$ and $f(t;\boldsymbol{\theta}_{\gamma}) = {\alpha}_{y,i} +  {\beta}_{y} t$, where both ${\alpha}_{x,i}$ and ${\alpha}_{y,i}$ are assumed to be normally distributed. 
Note that other parametrizations are also possible. More details on the choice of both functions is given in section \ref{sec:sccamm}.

\subsection{TOSCCA for Multiple Measurements} \label{sec:sccamm}

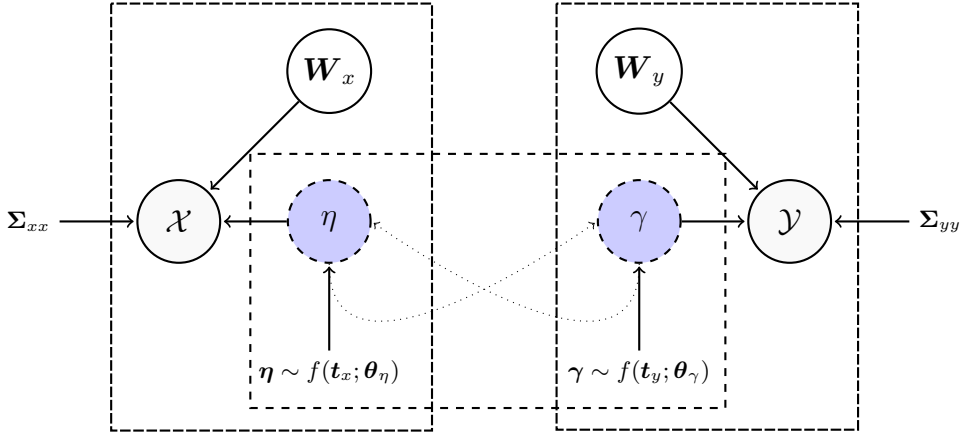
\begin{figure}[!h]
\centering
    \begin{tikzpicture}[
  shorten >=1pt,
  node distance=1.5cm and 1cm,
  auto,
  state/.style={circle, draw=black, thick, minimum size=1.1cm, font=\large, text centered, fill=blue!20},
  dashedstate/.style={circle, draw=black, dashed, thick, minimum size=1.1cm, font=\large, text centered, fill=blue!10},
  squarestate/.style={rectangle, draw=black, thick, minimum size=1.1cm, font=\large, text centered, fill=blue!10},
  label/.style={font=\small\itshape},
  every edge/.style={draw, thick}
]

  \node[] (1) {\color{white}$\mathrm{z}$};
  \node[dashedstate, fill=blue!20, above right of=1, xshift=1cm] (2) {$\gamma$};
  \node[dashedstate, fill=blue!20, above left of=1, xshift=-1cm] (3) {$\eta$};
  \node[state, fill=gray!5, right of=2, xshift=0.5cm] (4) {$\mathcal{Y}$};
  \node[state, fill=gray!5, left of=3, xshift=-0.5cm] (5) {$\mathcal{X}$};
  \node[state, fill=white!20, above of=2, yshift=0.5cm] (6) {$\bW_y$};
  \node[state, fill=white!20, above of=3, yshift=0.5cm] (7) {$\bW_x$};
  
  \node[label, below of=2, yshift=-0.5cm] (8) {$\bgamma \sim f(\bt_y;\boldsymbol{\theta}_{\gamma})$};
  \node[label, below of=3, yshift=-0.5cm] (9) {$\boldsymbol{\eta} \sim f(\bt_x;\boldsymbol{\theta}_{\eta})$};
  
  \node[label, right of=4, xshift=0.5cm] (10) {$\boldsymbol{\Sigma}_{yy}$};
  \node[label, left of=5, xshift=-0.5cm] (11) {$\boldsymbol{\Sigma}_{xx}$};

  \draw[->, dotted] (2.south) .. controls +(0,-2) and +(0,0) .. (3.east) node[pos=0.5, yshift=3.4ex] {};
  \draw[->, dotted] (3.south) .. controls +(0,-2) and +(0,0) .. (2.west) node[pos=0.5, yshift=3.4ex] {};

    
  \path[->, thick] (2) edge (4);
  \path[->, thick] (3) edge (5);
  \path[->, thick] (6) edge (4);
  \path[->, thick] (7) edge (5);
  \path[->, thick] (8) edge (2);
  \path[->, thick] (9) edge (3);
  \path[->, thick] (10) edge (4);
  \path[->, thick] (11) edge (5);


  \draw[thick, dashed] ($(3.north west)+(-0.65,0.5)$) rectangle ($(9.south east)+(4.2,-0.2)$);

  \draw[thick, dashed, dash pattern=on 4pt off 1pt] ($(3.north west)+(-2.5,2.5)$) rectangle ($(9.south east)+(0.3,-0.5)$);
  \draw[thick, dashed, dash pattern=on 4pt off 1pt] ($(4.north west)+(-2.7,2.5)$) rectangle ($(10.south east)+(-1.5,-2.5)$);

\end{tikzpicture}
    \caption{Illustration of TOSCCA-MM. {Diagram of the Thresholded Ordered Sparse Canonical Correlation Analysis for Multiple Measurements (TOSCCA-MM) model. Observation matrices $\mathcal{X}$ and $\mathcal{Y}$ are in shaded nodes. Latent variables $\boldsymbol{\eta}$ and $\bgamma$, in blue-shaded dashed nodes, capture the correlation between them through sparse canonical vector $\bW_x$ and $\bW_y$. We model the longitudinal dimension in $\boldsymbol{\eta}$ over $\bt_x$ for each latent variable, which implicitly corrects for the correlation between measurements from the same individual. Similarly, the longitudinal dimension in $\bgamma$ over $\bt_y$ is modelled through $f_{\bt}$. See steps \ref{st:eta_estimate} and \ref{st:gamma_estimate} of Algorithm \ref{alg:scca} for more detail. We then use the estimated model to calculate predicted values for $\boldsymbol{\eta}$ (and similarly $\bgamma$) over $\bt_y$ (and similarly $\bt_x$), steps \ref{st:eta_pred} and \ref{st:gamma_pred}. The dotted arrows indicate how these values are used to obtain the respective canonical weights, steps \ref{step:beta_eta} and \ref{setep:alpha_gamma}. The residual covariance matrices, $\boldsymbol{\Sigma}_{xx}$ and $\boldsymbol{\Sigma}_{yy}$, are made up from residual noise covariances and variation specific to each dataset.}}
    \label{fig:diag_toscamm}
\end{figure}

We refer to our extended model as multiple measurements TOSCCA-MM. The algorithm is described in pseudo-code in Algorithm \ref{alg:scca}. The steps may be divided into two separate sections: finding sparse canonical weights $\bW_x$ and $\bW_y$, steps \ref{step:beta_eta}-\ref{st:soft_beta} and \ref{setep:alpha_gamma}-\ref{st:soft_alpha} (described in section \ref{sec:toscca}) 
and modelling the latent trajectories, steps \ref{st:gamma_estimate}-\ref{st:gamma_pred} and \ref{st:eta_estimate}-\ref{st:eta_pred} 
(section \ref{sec:sccamm}). Note that the original TOSCCA algorithm for non-longitudinal problems can be obtained by removing steps \ref{st:eta_estimate}, \ref{st:eta_pred}, \ref{st:gamma_estimate} and \ref{st:gamma_pred} of Algorithm \ref{alg:scca}. Figure \ref{fig:diag_toscamm} illustrates the concept behind TOSCCA-MM.

We address the correlation due to multiple measurements in the latent space by modelling the time dependency in $\boldsymbol{\eta}_i$ and $\bgamma_i$ for all $\bt_{x,i}$ and $\bt_{y,i}$, respectively, as in equations \eqref{eq:obs_lv_long_g} and \eqref{eq:obs_lv_long_e}. After fitting model $f(\bt_{x,i};\boldsymbol{\theta}_{\eta})$ over $\boldsymbol{\eta}_i$ and $f(\bt_{y,i};\boldsymbol{\theta}_{\gamma})$ over $\bgamma_i$, we predict $\boldsymbol{\eta}_i$ for $\bt_{y,i}$ and $\bgamma_i$ for $\bt_{x,i}$ to overcome disparities between the measurements of each dataset. Given the latent variables, this approach yields conditional independence, due to the structure of the NIPALS algorithm. In doing so, we gain computational efficiency by dealing with said correlations in the low dimensional space. Additionally, this approach renders latent paths for all measurements of each sample. These paths represent the changing underlying trajectories for the correlated components of each dataset. After estimating our canonical weights in the $l^{th}$ iteration, we are left with a latent variables $\bgamma$ and $\boldsymbol{\eta}$ of $n$ samples with at least one measurement per sample $i$. These latent variables describe a single latent mechanism linking both datasets, for each component $k$.

\begin{algorithm}
\footnotesize
    \caption{TOSCCA-MM}\label{alg:scca}
    \textbf{Input. } $\mathcal{X}$, $\mathcal{Y}$, $\bW_x^{(0)}$, $p_{x}$, and $q_{y}$\\
    \textbf{Output. } $\bw_{x,k}^*$ and $\bw_{y,k}^*$ \\
    $l \gets 1$, $\xi << 1$, $\varepsilon = 10^6$, $\rho^{(0)} \gets 0$
    \begin{algorithmic}[1]
    \While{$\varepsilon > \xi$} \Comment{Changes larger than tolerance measure $\xi$}
    \State $\boldsymbol{\eta} \gets \mathcal{X}\bw_x^{(l-1)}$ \label{st:eta0}
    \State $\hat{\boldsymbol{\eta}}_{i} \gets \hat{\alpha}_{x,i} + \hat{\beta}_{x}\boldsymbol{t}_{x,i}$ \Comment{Estimate $f_x$, which is a LME model with random intercepts and fixed slope} \label{st:eta_estimate}
    \State $\tilde{\boldsymbol{\eta}}_{i} \gets \hat{\alpha}_{x,i} + \hat{\beta}_{x}\boldsymbol{t}_{y,i}$ \Comment{Get predicted values for measurements in $\boldsymbol{t}_{y,i}$} \label{st:eta_pred}
    \State $\tilde{\bw}_y^{(l)}\gets \mathcal{Y}^{\intercal}\tilde{\boldsymbol{\eta}}$ \label{step:beta_eta}
    \State $\bw_{y,k}^{(l)} \gets \mathds{1}_{\scaleto{|\tilde{\bw}_y^{(l)}|>q_{y}}{6pt}}\tilde{\bw}_y^{(l)}  - q_{y}$ \label{st:soft_beta}
    \State $\bgamma_k \gets \mathcal{Y}\bw_{y,k}^{(l)}$ \Comment{Standardise canonical variable for $\mathcal{Y}$} \label{st:gamma}
    \State $\hat{\bgamma}_i \gets\hat{\alpha}_{y,i} + \hat{\beta}_{y}\boldsymbol{t}_{y,i}$ \Comment{Equivalent to step \ref{st:eta_estimate}} \label{st:gamma_estimate}
    \State $\tilde{\bgamma}_i \gets\hat{\alpha}_{y,i} + \hat{\beta}_{y}\boldsymbol{t}_{x,i}$ \Comment{Get predicted values for measurements in $\boldsymbol{t}_{x,i}$} \label{st:gamma_pred}
    \State $\tilde{\bw}_x^{(l)} \gets \mathcal{X}^{\intercal}\tilde{\bgamma}_k$ \label{setep:alpha_gamma}
    \State $\bw_{x,k}^{(l)} \gets \mathds{1}_{\scaleto{|\tilde{\bw}_x^{(l)}|>p_{x}}{6pt}}\tilde{\bw}_x^{(l)} - p_{x}$ \label{st:soft_alpha}
    \State $\boldsymbol{\eta}_k \gets \mathcal{X}\bw_{x,k}^{(l)}$ \Comment{Standardise canonical variable for $\mathcal{X}$}
    \State $\rho^{(l)} \gets cor(\boldsymbol{\eta}_k, \bgamma_k)$
    \State $\varepsilon \gets \rho^{(l)} - \rho^{(l-1)}$ 
    \State $l = l+1$
\EndWhile\label{sccaWhile}
\State \textbf{return} $(\bw_{x,k}^*, \bw_{y,k}^*)$\Comment{Canonical weights}
\end{algorithmic}
\end{algorithm}

The choice of $f(\bt;\cdot)$ will influence the composition of the components, choosing which dynamic is represented in the model. These may yield different interpretations or representations of the underlying mechanisms in the data. Using the example in section \ref{sec:tim_dyn}, in step \ref{st:eta_estimate} of Algorithm \ref{alg:scca} we have 
$f(\bt_x\boldsymbol{\hat{\theta}_{\eta}}) = \hat{\alpha}_{x,i} +  \hat{\beta}_{x}\boldsymbol{t}_{x,i}$, which estimates a LME with random intercepts and fixed slope for the latent variable. We then use the estimated parameters $\boldsymbol{\hat{\theta}}_{\eta} = \{\hat{\alpha}_{x,i},\hat{\beta}_{x}\}$ to predict the latent values over the time vector $\bt_y$ (step \ref{st:eta_pred}), $\tilde{\boldsymbol{\eta}} = f(\bt_y; \boldsymbol{\hat{\theta}}_{\eta})$. That is, our latent variables are expectations at times $\bt_x$ and $\bt_y$ of the longitudinal model specified for $f(\cdot)$. 
Finally, passing on these predicted latent values to estimate the alternate canonical vectors $\bW_y$, as $\bW_y^{(l+1)} = \left(\bY^T\bY\right)^{-1}\bY^T \tilde{{\bgamma}}^{(l)}$, at each iteration $l$. These predicted values fill in the blanks due to sparse and irregularly observed data. 
Likewise, for step \ref{st:gamma_estimate} we have $f(\bt_y;\boldsymbol{\hat{\theta}_{\gamma}}) = \hat{\alpha}_{y,i} +  \hat{\beta}_{y}\boldsymbol{t}_{y,i}$, where $\boldsymbol{\hat{\theta}}_{\gamma} = \{\hat{\alpha}_{y,i},\hat{\beta}_{y}\}$. We get the predicted values over $\bt_x$ in step \ref{st:gamma_pred}.

Through this approach, we correct for the correlated measurements ($Cov(\bx_{i,t}, \bx_{i,v}) \neq \boldsymbol{0} \; \text{for} \; t \neq v, \, (t,v)\in t_{x,i}$) relying on the criss-cross nature of the NIPALS algorithm which removed said correlation conditional on the latent variable, assuming this one is completely described, $Cov(\bx_{i,t}, \bx_{i,v}|\bgamma_{i,t})  = \boldsymbol{0}$.

Modelling the latent variables longitudinally reveals trajectories that can be associated with the dynamics of underlying mechanisms connecting the datasets. Additionally, because the latent variables are low-dimensional, this approach efficiently handles the correlated measurements in high-dimensional CCA. Finally, the function $f(\cdot)$ offers flexibility, allowing for the representation of various theories about the latent paths.

\section{Results}

\subsection{Simulations}\label{sec:sim}

\begin{figure}[!h]
    \centering
    \begin{subfigure}{0.45\textwidth}
        \centering
        \includegraphics[width=\linewidth]{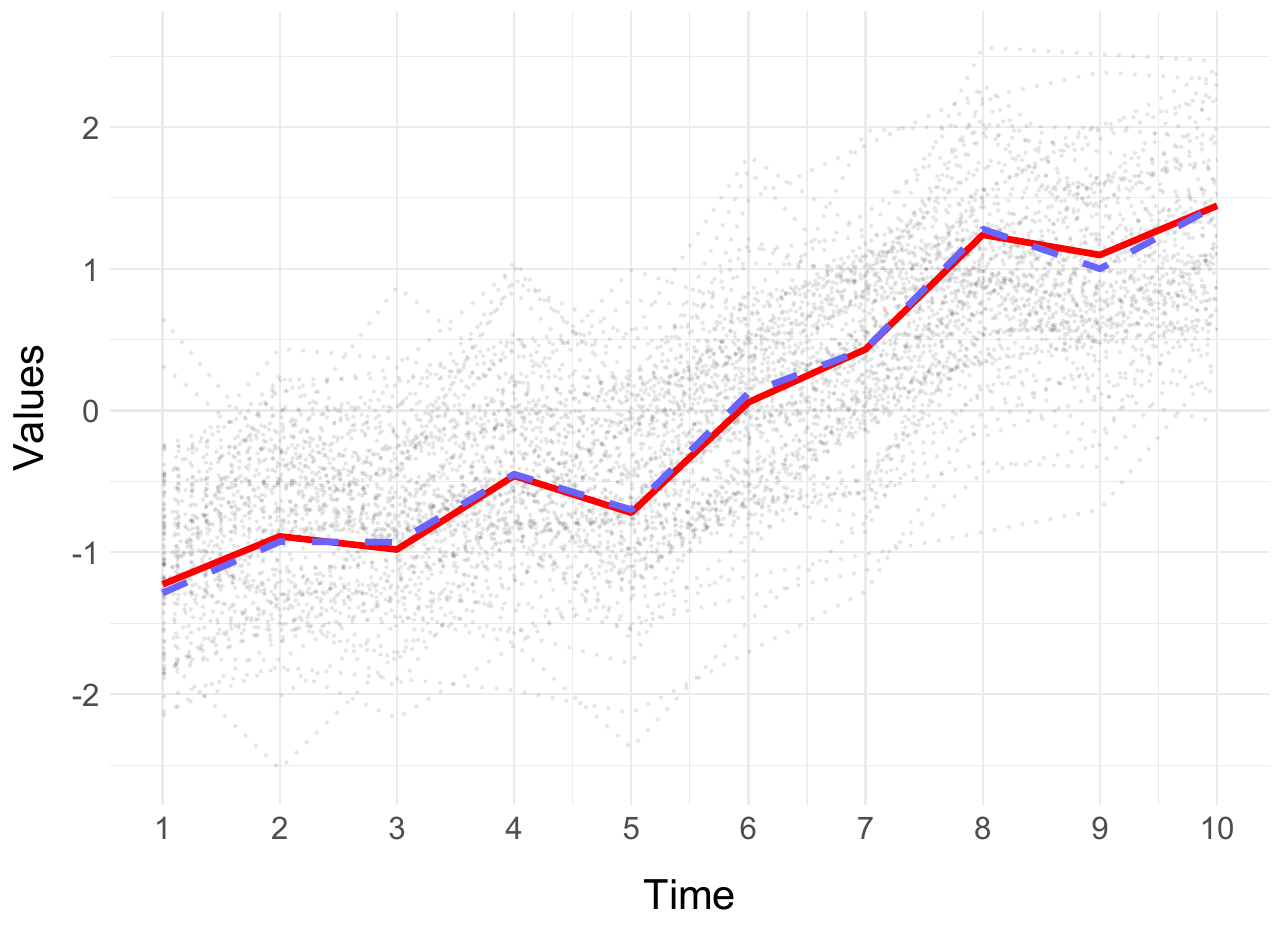} 
        \caption{First component $k=1$}
        \label{fig:sim_k1}
    \end{subfigure}
    \hfill
    \begin{subfigure}{0.45\textwidth}
        \centering
        \includegraphics[width=\linewidth]{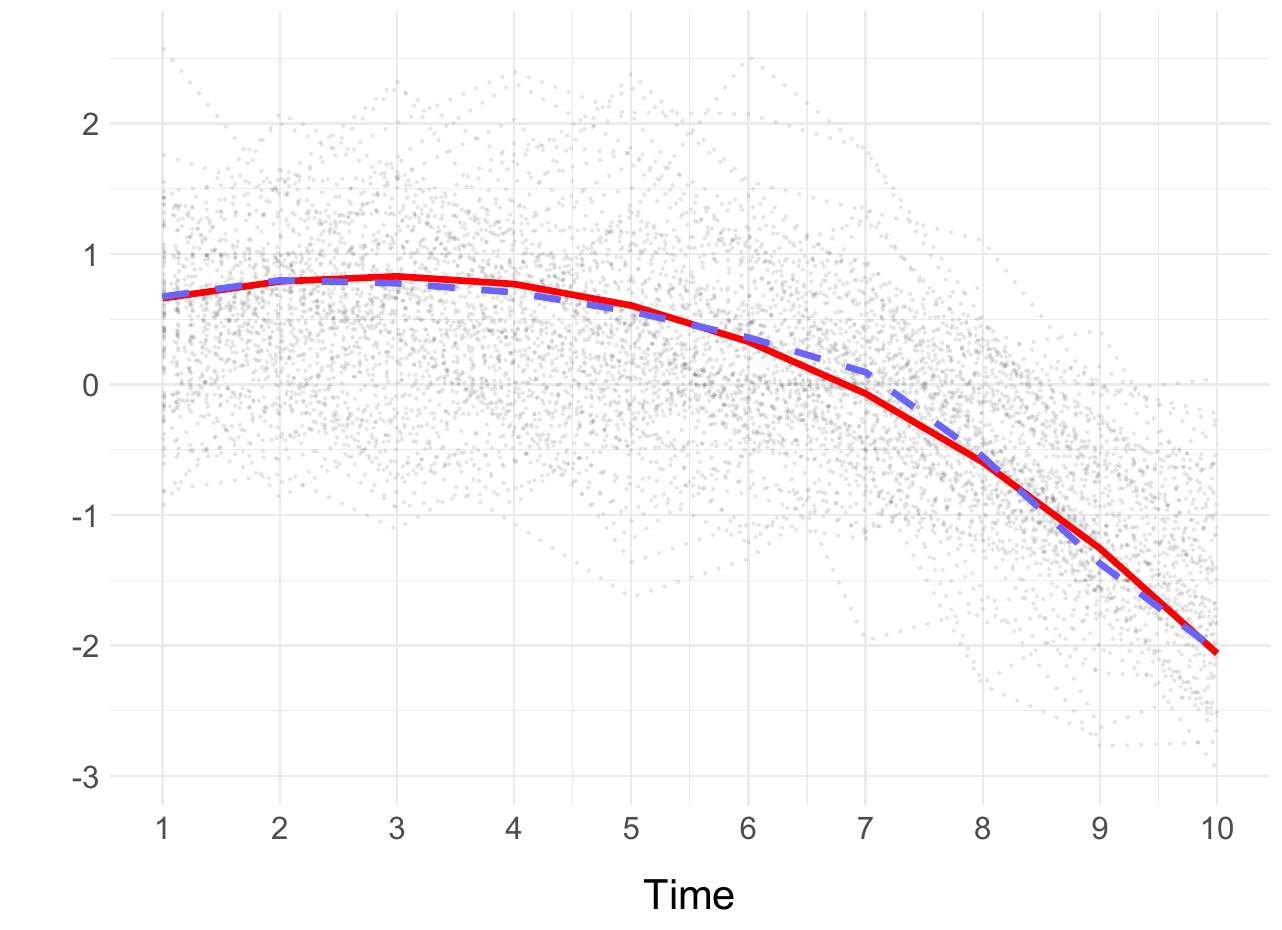} 
        \caption{Second component $k=2$}
        \label{fig:sim_k2}
    \end{subfigure}
    
    \vspace{0.5cm} 
    \begin{subfigure}{0.8\textwidth}
        \centering
        \includegraphics[width=0.65\linewidth]{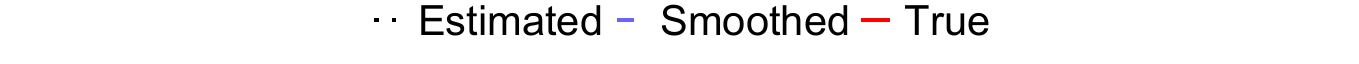} 
    \end{subfigure}
    
    \caption{Estimated and true latent trajectories of components $k=1$ and $k=2$}
    \label{fig:sim_lv}
\end{figure}

We simulated our data based of the probabilistic generating process described by Bach and Jordan (2005) \parencite{bach_probabilistic_2005}, where instead of two latent variables, we only assume one, $\bz$. For each measurement $m = [1, 10]$, we simulated data for $n=100$ samples with $p=10000$ and $q=200$ variables in $\mathcal{X}$ and $\mathcal{Y}$, respectively. Then, we artificially removed at random $20\%$ measurements from $\mathcal{X}$ and $30\%$ from $\mathcal{Y}$. The same individual could have missing none, multiple or even all of their measurements. 
That is, for each point in time $t\in\bt_x$ and sample $i\in [1, n]$ we generate our data following equation \eqref{eq:sim_gen}:

\begin{equation}\label{eq:sim_gen}
    \bx_{i} \sim \mathcal{N}(z_{i}\bW_x^T, \boldsymbol{\Psi}_x), 
\end{equation}

where the latent variable $\bz$ has measurements for all times in $\bt$ and $\boldsymbol{\Psi}_x$ is a $\bar{t}\times \bar{t}$ matrix, where $\bar{t} = \text{length}(\bt)$, with ones on the main diagonal and some off-diagonal nonzeros to model time dependence. 

We simulated the latent trajectories for the first component as $\bz_{1,i} = \theta_0\bt + \sin(\theta_1 \bt)^{\bt} + \boldsymbol{\epsilon}_{1,i}$ and $\bz_{2,i} = \theta_2\bt + \left(1 + \frac{\bt}{\max(\bt)}\right)^3 + \boldsymbol{\epsilon}_{2,i}$ for the second component.
We followed the same scheme to generate the values for $\mathcal{Y}$. 

We used the TOSCCA-MM algorithm to find $\bW_x$ and $\bW_y$ over all measurements and samples and draw the paths in the latent space. To recover the latent paths for both components ($k=1$ and $k=2$), we chose a mixed effect model $z_{k,i} \sim \theta_{0,i} + P_3(t_{x,i}, \boldsymbol{\theta}_1)$, where  $P_3(t_{x,i}, \boldsymbol{\theta}_1)$ is a third degree polynomial with fixed effects vector $\btheta_1$. The latent paths were successfully recovered, as shown in figures \ref{fig:sim_k1} and \ref{fig:sim_k2}, where the black dotted lines represent estimated individual trajectories, the dashed blue line is the estimated average of said trajectories, and the solid red line is the true latent path. Furthermore, TOSCCA-MM correctly included most relevant weights, with the exception of the canonical weights for $\mathcal{X}$, $\bw_{x,1}$, in which, out of $10$ nonzero weights and TOSCCA-MM found $5$; and, even though some weights that should have been zero, were estimated to be different from zero, most of them remained around zero. Last, we calculated additional components to ensure TOSCCA-MM returned \textit{empty} latent paths (figure \ref{fig:sim_k3} section \ref{app:coeffs}).

\begin{figure}[!h]
    \centering
    \includegraphics[width=0.65\linewidth]{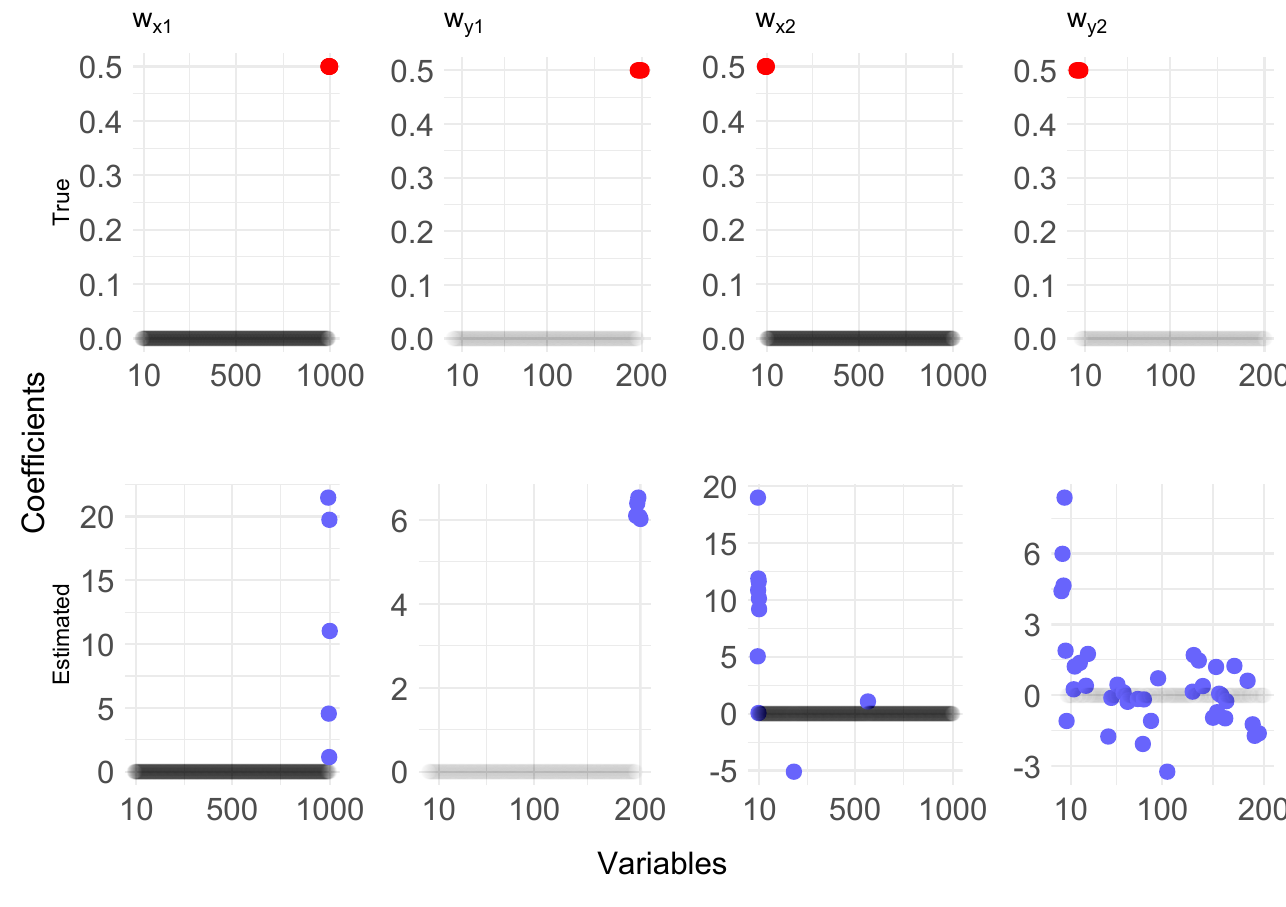}
    \caption{True ({\color{red}red}) and estimated (\textcolor{blue!60}{blue}) canonical weights.}
    \label{fig:can_weights}
\end{figure}

\subsection{HMP data}\label{sec:hmp}
We used data from the Human Microbiome Project (HMP) on gene expression measurements and gut microbiome signatures from 84 healthy and prediabetic individuals \parencite{hmp_zhou_2019} and looked into the gut-gene associations that may differ between insulin sensitive (IS) and insulin resistant (IR) individuals.

\begin{figure}
    \centering
    \includegraphics[width=1\linewidth]{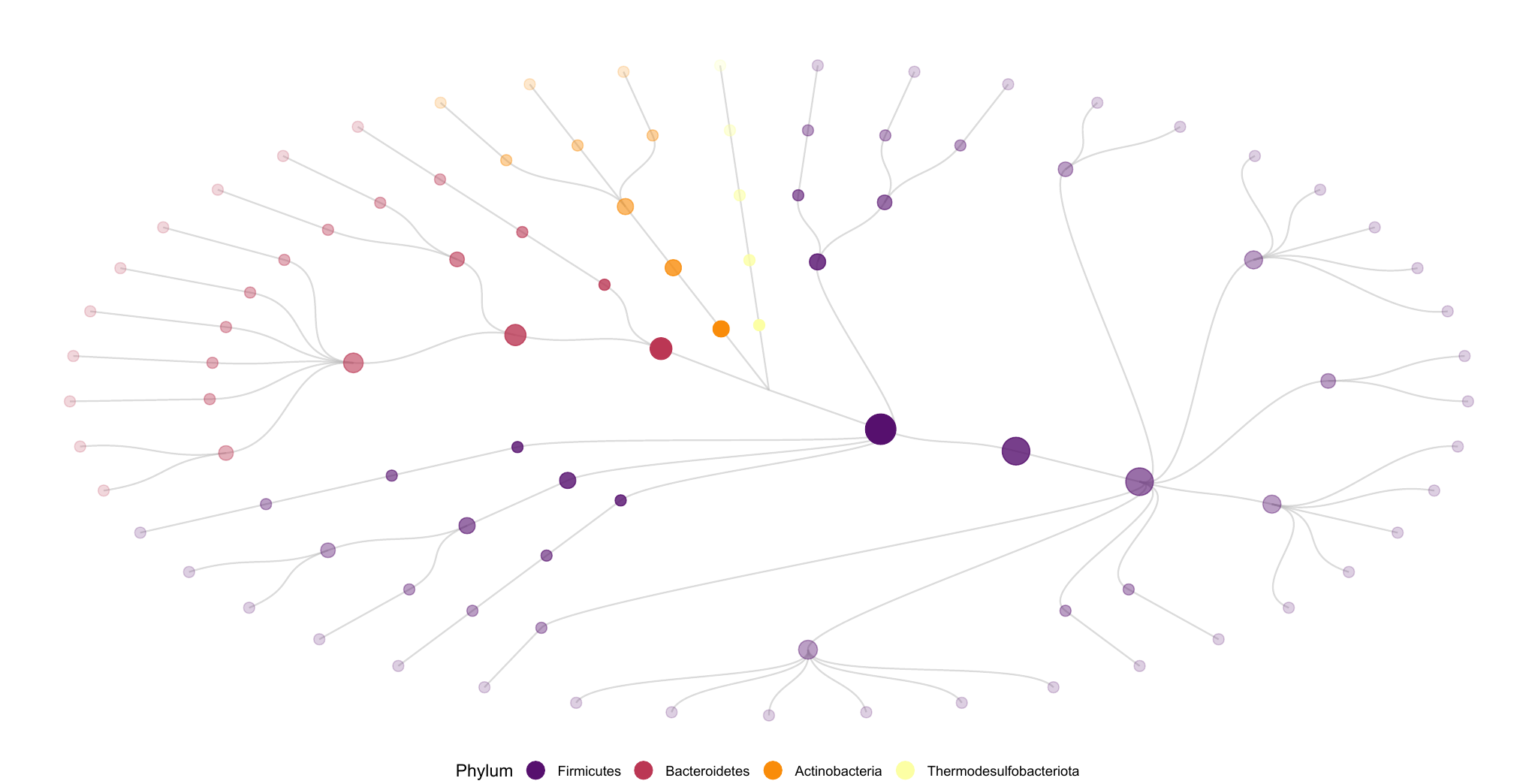}
    \caption{Taxonomic tree}
    \label{fig:tree}
\end{figure}

The gut microbiome in our data is divided into four big groups: Firmicutes, Bacteroidetes, Actinobacteria and Thermodesulfobacteriota. The first two make up 90\% of humans gut microbiota. These groups are organised into five hierarchical aggregation levels: genus, family, order, class, and phylum, with phylum representing the highest level of aggregation. We show in Figure \ref{fig:tree} these groups by colour and the levels by shades. That is, the nodes on the outer side with the lightest shade represent Operational Taxonomic Units (OTUs) at the genus level. Each step into the centre inwards, aggregates the previous levels.

\subsubsection{Gut-gene associations and prediabetes}
Prediabetes is a condition arising from insufficient production of insulin in the pancreas to maintain normal blood glucose levels. It is considered a precursor to diabetes, as individuals with prediabetes have elevated blood glucose levels, though below those of diabetic patients. Most people with prediabetes experience IR and many with IR eventually develop prediabetes. 

In the past decade, studies have widened the scope of mechanisms involved in the development of (pre-)diabetes and diabetes. Inflammatory immune responses have been linked to the development of prediabetes by contributing to IR. Immune and metabolic disruptions influence and are influenced by the gut microbiome. The gut microbiome is a balanced ecosystem unique to each individual affecting multiple layers of human health. Among those layers, the composition of the gut microbiota plays an important role in the continuous development of the immune system with ties to risks related to the development of prediabetes and type 2 diabetes. Alterations in said composition are connected to IR \parencite{prediabGut_wu_2020, gutIR_takeuchi_2023, gutInflammation_semo_2024}. Particularly, the abundance of some microbes linked to inflammation, such as \textit{Prevotella} and some groups of \textit{Clostridiales} and \textit{Actinobacteria}, has been used to predict risk of prediabetes \parencite{predGut3_pinna_2021, predGut3_alvarezSilva_2021}. These alterations may be triggered by health disruptions as healthy individuals have different responses compared to their insuline sensitive (IS) counterparts \parencite{hmp_zhou_2019}.

A host of genetic factors are involved in the make up of the gut microbiota as well as influence its alterations\parencite{geneGut_nichols_2020}. Genome Wide Association Studies (GWAS) have found that genetic variants could alter gene expression patterns which in turn may have an effect on the intensity of the immune response \parencite{geneGut_nichols_2020, geneImmune_diedisheim_2020}. Additionally, IR patients show differences in their gene expression patterns related to key enzymes responsible for processing glucose \parencite{gutGeneGlu_patti_2004}.

\begin{figure}
    \centering
    \includegraphics[width=0.9\linewidth]{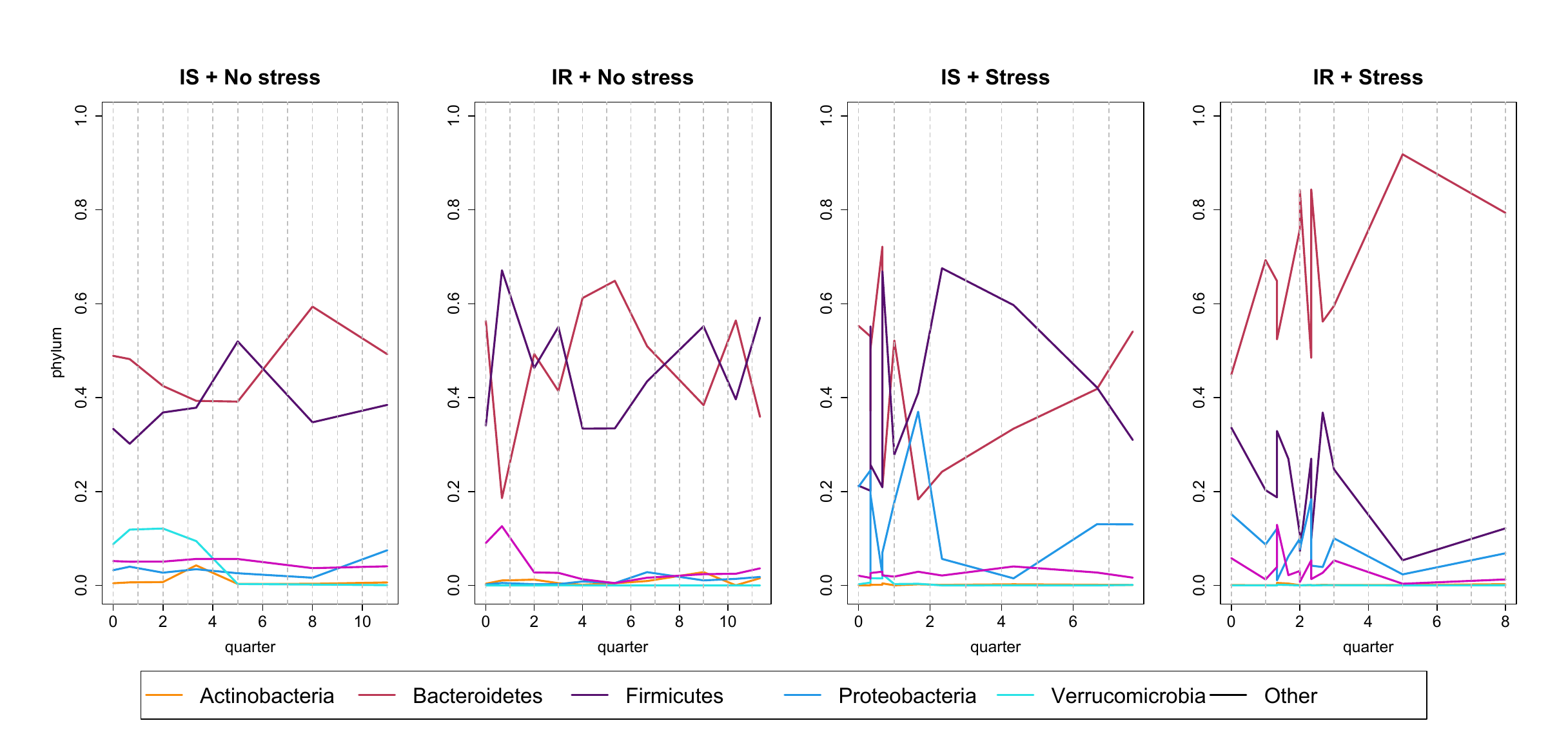}
    \caption{OTUs measurements at phylum level for four patients.}
    \label{fig:example_first}
\end{figure}

We observed this pattern in our data looking at the dynamics of the OTUs at the phylum level for patients who were IS, IR and with or without a a negative health event during the study, for which data is available. We referred to these visits as \textit{stress} visits, as opposed to healthy, regular visits. Figure \ref{fig:example_first} shows examples for all four scenarios. Patients with no additional \textit{stress} to their health had more stable OTU values than those who did experienced it, while IS and IR patient appeared to be affected by these differently.

\subsubsection{Results from TOSCCA-MM analysis of HMP Data}
We used measurements on over 10000 genes and 21 OTUs at the family level.\footnote{Results for the other taxonomic levels are left to Appendix \ref{app:hmp_levels} for brevity.} Each individual had between 1 and 80 measurements approximately every three months. As mentioned in section \ref{sec:intro}, these were irregularly and sparsely observed. That is, visits were not recorded at all times for every individual, or there were multiple visits within those three months. The latter occurred when an important event in the patient's health was detected (usually via self report). The most common of these events were falling sick (and taking medication) and weight changes. These were all categorised as \textit{stress} visits. We kept the first 19 measurements for each view, after which data was available on very few patients. 

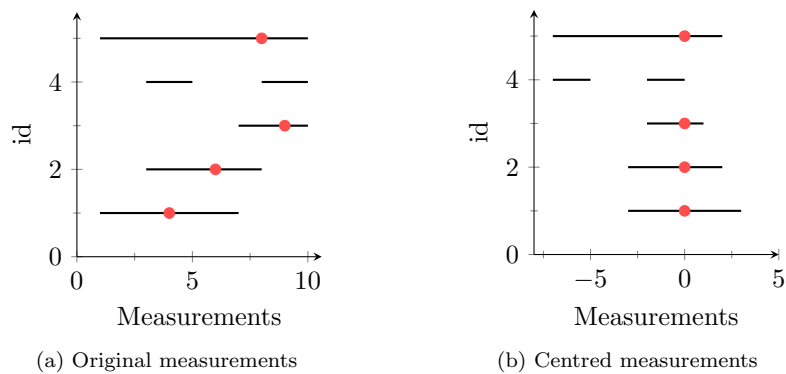
\begin{figure}[h]
    \centering
    \begin{subfigure}{0.32\linewidth}
        \centering
        \begin{tikzpicture}
            \begin{axis}[
                width=1\linewidth,
                height=1\linewidth,
                xlabel={Measurements},
                ylabel={id},
                ymin=0, ymax=5.6,
                xmin=0, xmax=10.6,
                axis lines=left,
                minor tick num=1,
                legend pos=south east
            ]
                \addplot[domain=1:7, samples=2, color=black, thick] {1} node[pos=0.1, anchor=south west] {};            
                \addplot[domain=3:8, samples=2, color=black, thick] {2} node[pos=0.1, anchor=south west] {};
                \addplot[domain=7:10, samples=2, color=black, thick] {3} node[pos=0.1, anchor=south west] {};
                \addplot[domain=3:5, samples=2, color=black, thick] {4};
                \addplot[domain=8:10, samples=2, color=black, thick] {4} node[pos=0.1, anchor=south west] {};
                \addplot[domain=1:10, samples=2, color=black, thick] {5} node[pos=0.1, anchor=south west] {};

                \addplot[only marks, mark=*, mark options={color=red!70}] coordinates {(8,5)};
                \addplot[only marks, mark=*, mark options={color=red!70}] coordinates {(9,3)};
                \addplot[only marks, mark=*, mark options={color=red!70}] coordinates {(6,2)};
                \addplot[only marks, mark=*, mark options={color=red!70}] coordinates {(4,1)};
            \end{axis}
        \end{tikzpicture}
        \caption{Original measurements}
        \label{fig:hmp_measurement_left}
    \end{subfigure}
    \hspace{3em}
    \begin{subfigure}{0.32\linewidth}
        \centering
        \begin{tikzpicture}
            \begin{axis}[
                width=1\linewidth,
                height=1\linewidth,
                xlabel={Measurements},
                ylabel={id},
                ymin=0, ymax=5.6,
                xmin=-8, xmax=5,
                axis lines=left,
                minor tick num=1,
                legend pos=south east
            ]
                \addplot[domain=-3:3, samples=2, color=black, thick] {1} node[pos=0.1, anchor=south west] {};            
                \addplot[domain=-3:2, samples=2, color=black, thick] {2} node[pos=0.1, anchor=south west] {};
                \addplot[domain=-2:1, samples=2, color=black, thick] {3} node[pos=0.1, anchor=south west] {};
                \addplot[domain=-7:-5, samples=2, color=black, thick] {4};
                \addplot[domain=-2:0, samples=2, color=black, thick] {4} node[pos=0.1, anchor=south west] {};
                \addplot[domain=-7:2, samples=2, color=black, thick] {5} node[pos=0.1, anchor=south west] {};

                \addplot[only marks, mark=*, mark options={color=red!70}] coordinates {(0,5)};
                \addplot[only marks, mark=*, mark options={color=red!70}] coordinates {(0,3)};
                \addplot[only marks, mark=*, mark options={color=red!70}] coordinates {(0,2)};
                \addplot[only marks, mark=*, mark options={color=red!70}] coordinates {(0,1)};
            \end{axis}
        \end{tikzpicture}
        \caption{Centred measurements}
        \label{fig:hmp_measurement_right}
    \end{subfigure}
    \caption{Original and centred measurements around an event (\textcolor{red!70}{red dot}).}
    \label{fig:hmp_measurements}
\end{figure}

We rearranged the longitudinal dimension by summarising visits into three month intervals centred around the first \textit{stress} visit (Figure \ref{fig:hmp_measurements}). Measurements from patients without a reported \textit{stress} visit were shifted to the left (Figure \ref{fig:hmp_measurement_left}). These changes turned our multiple measurements into comparable time-series among all individuals and allowed us to observe the latent mechanisms around health events which are known to follow different mechanisms between healthy and glucose-dysregulated people. Last, we divided the observations into two groups, insulin-sensitive (IS) and IR, and performed the analysis on each group.

We used the TOSCCA-MM algorithm algorithm to find the canonical vectors and fit their latent trajectories. We found the optimal threshold parameters $p_{\alpha}$ and $q_{\beta}$  via cross-validation over a grid of possible combinations. Because we centred our time-series along a particular event, we chose a change-point mixed effect model with random intercepts and a polynomial slope (equation \eqref{eq:hmp_latent_poly}). Different choices of parametrisation for equation \eqref{eq:hmp_latent_poly} would of course, render different compositions or change the order in which the components appeared.\footnote{For this case, we compared alternative specifications for the latent trajectories and observed virtually equivalent components with minor changes at some of the taxonomic levels.} The specification in equation above was general enough to find trend, inflection or/and change points with random intercepts.

\begin{align} \label{eq:hmp_latent_poly}
\gamma_{i,t} = a_{0,i}+ \sum_{r}^3a_{r}t^r + \beta t\mathbbm{1}_{t>s}+ \varepsilon_{i,t}
\end{align}

\begin{figure}[!h]
    \centering
    \includegraphics[width=0.72\linewidth]{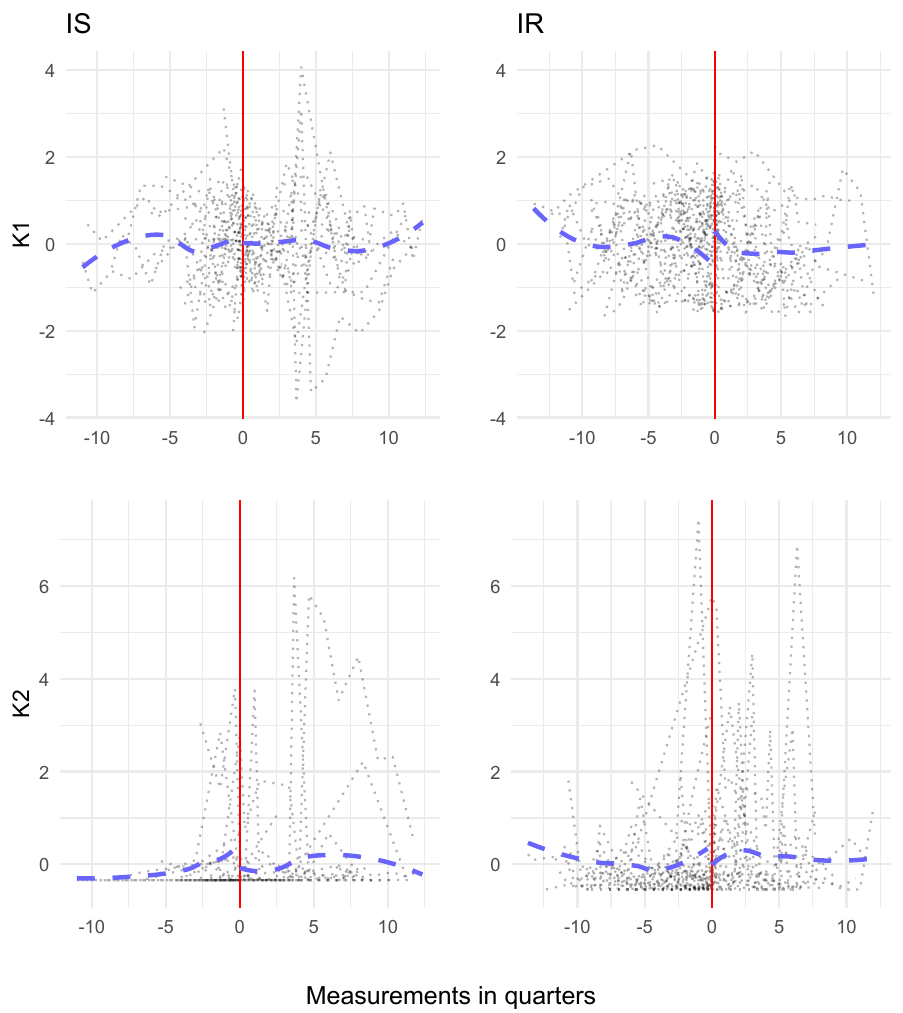}
    \caption{Estimated latent paths for component $k=1$ and $k=2$.}
    \label{fig:hmp_lv_fam}
\end{figure}

The latent trajectories for the first two components are plotted in Figure \ref{fig:hmp_lv_fam}. The first component was made up from the most abundant OTUS for both groups and appeared to show similar latent trajectories representing general links between the gut microbiota and the host gene expressions. Hence, picking up on the general conditions of the gut microbiome for both groups, with their latent trajectories selecting the same OTUs. Additionally, the path for the IR group showed a disturbance at the at the moment of the health event compared to the healthy group. This could lead back to IR patients experiencing grater disruptions than their healthy counterparts in such episodes. 
On the other hand, the second component was composed of different OTUs for each group. The latent mechanism for the IS group was made up of \textit{Prevotellaceae}, linked to inflammatory events, generally responding to altered immune and gut environments an insulin regulation.
For the IR group, \textit{Clostridiales} and \textit{Coriobacteriaceae} were the relevant OTUs. The latter is associated to insulin sensitivity mechanism through its production of \textit{butyric acid}.  \textit{Coriobacteriaceae} have been linked to benefits in insulin sensitivity in the prevention of the development of Type 2 Diabetes. The second component appeared to put forward differences between groups with the IR group giving relevance to OTUs important for insulin regulation.

The above may indicate that the first component represented the trivial association between genetics and gut composition; while the second indeed described differences between IS and IR trajectories under stress. Thus, suggesting relevant latent dynamics are different between these groups. 

With respect genetic contributors to each mechanism we observed a similar pattern in the first component. The main contributors were shared between both IS and IR individuals comprising of around 200 estimated nonzero canonical weights. For the second component, however, the estimated nonzero weights were not the same. This once more indicated differences between the groups on the dynamics which we previously linked to the \textit{stress} events, on the genetic level.

Whether these differences in component composition stem from reactions associated with their IS or IR status, or from tendencies toward distinct health disruptions within each group, falls outside the scope of this paper.

\section{Conclusions}

In this paper, we introduced a novel method, TOSCCA-MM, for building latent trajectories linking two datasets two high-dimensional datasets over time. Our approach captures the evolving dynamics of associations between variables by keeping the sparse canonical weights fixed across all measurements. TOSCCA-MM contrasts with existing methods which typically focus on correcting the correlations between measurements or estimating canonical weights at each point in time. These methods struggle with high-dimensions, sparse and irregularly observed data or fail to exploit the longitudinal aspect of the repeated measurements. By fixing the canonical weights, TOSCCA-MM prioritises uncovering the underlying processes driving the relationships between variables as they change together over time, rather than merely describing static correlations at each measurement. This enables us to draw latent trajectories of the most relevant features for each component, providing deeper insights into the hidden, time-dependent links between observation matrices.

TOSCCA-MM identifies sets of variables that are consistently highly correlated across measurements, and models how these correlations evolve over time in the latent space. This approach highlights key contributors to the latent dynamics and enables modelling of more complex patterns, such as lagged effects or variables that switch on and off over time. This features are directly built into TOSCCA-MM, making it more robust and interpretable compared to existing methods, which often struggle with the dimensionality of the data and the complexity of longitudinal relationships. 

TOSCCA-MM is particularly well-suited for handling high-dimensional data with irregular and sparsely observed data.
The implementation of TOSCCA-MM is available as open-source code on GitHub (\url{https://github.com/nuria-sv/toscca-mm}).

\printbibliography


\newpage
\appendix
\section{Appendix}

\subsection{Simulations} \label{app:coeffs}

\begin{figure}[!h]
    \centering
    \includegraphics[width=0.65\linewidth]{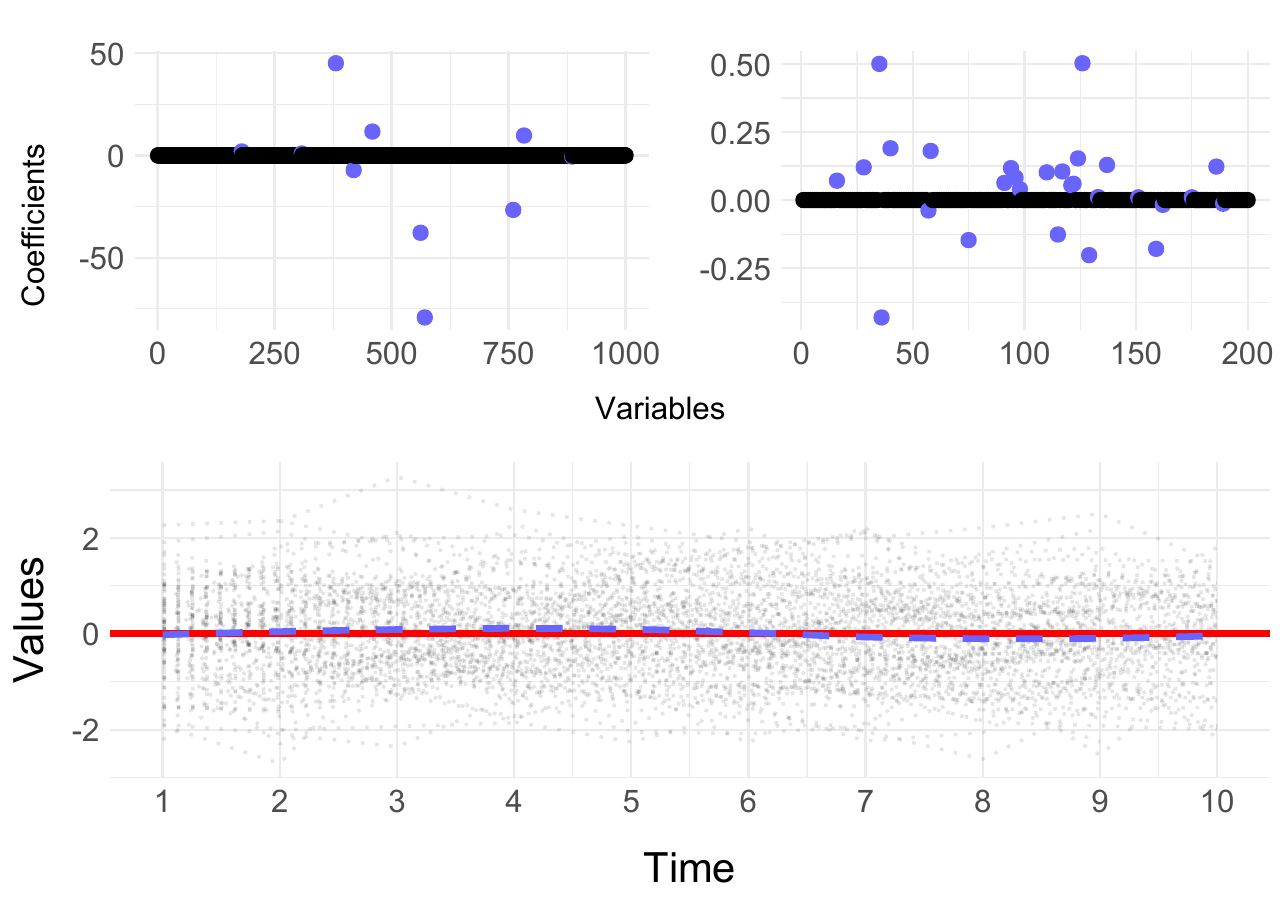}
    \caption{Estimated coefficients and latent path for $k=3$ (no signal).}
    \label{fig:sim_k3}
\end{figure}

Figure \ref{fig:sim_k3} displays the estimated coefficients and latent path for $k=3$ for which there is no signal. TOSCCA-mm selected some coefficients to be nonzero, most of them with values close to zero, with some extreme exceptions. The latent variable is correctly set to zero for all possible measurements. 

\subsection{HMP data} \label{app:hmp_levels}
Below we provide a quick summary of the estimated canonical vectors for the OTU's at the remaining levels: from most disaggregated (genus) to most aggregated (phylum). Different aggregation levels may put forward different dynamics hence we did not expect all estimates to perfectly align. In fact, some middle aggregation levels may mas smaller variations from different bacteria that may be visible at lower levels. On the other hand, higher levels may pick up on global effects for the different mayor microbiome groups. 

For all levels, including at the family level in the main text, the first component ($k=1$) highlighted general contributions and dynamics in both IS and IR patients. That is, the canonical vectors for both groups had virtually identical contributors for the first latent path. For that, in this section we focused on the second component ($k=2$) for which groups exhibited clear differences. 

\subsubsection{Genus}
\begin{figure}[!htbp]
    \centering
    \includegraphics[width=0.75\linewidth]{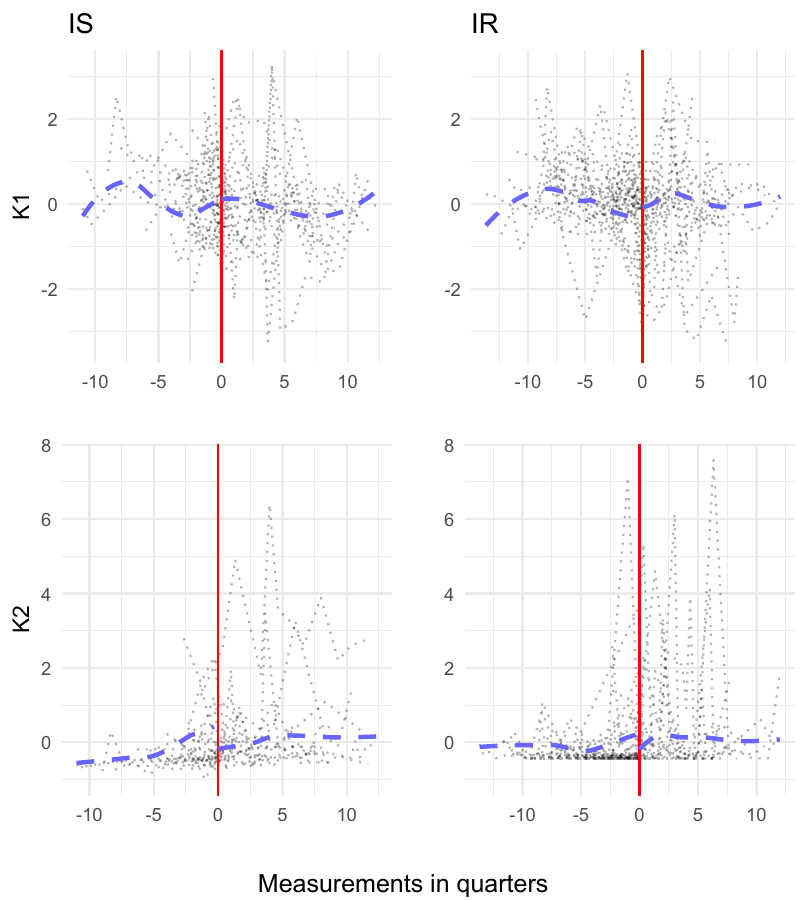}
    \caption{Estimated latent paths for $k=1$ and $k=2$ by group at the genus level.}
    \label{fig:hmp-genus}
\end{figure}

The estimated latent paths at genus level are displayed in figure \ref{fig:hmp-genus}. The canonical weights at the genus level for $k=2$ are not dissimilar to those discuss in the main text. 
The second component highlights OTU's linked to insulin sensitivity and anti-inflammatory systems, for the IS group. For the IR group, the only OTU for component $k=2$ \textit{Clostridiales Incertae Sedis XIII}.

\subsubsection{Order}
\begin{figure}[!htbp]
    \centering
    \includegraphics[width=0.75\linewidth]{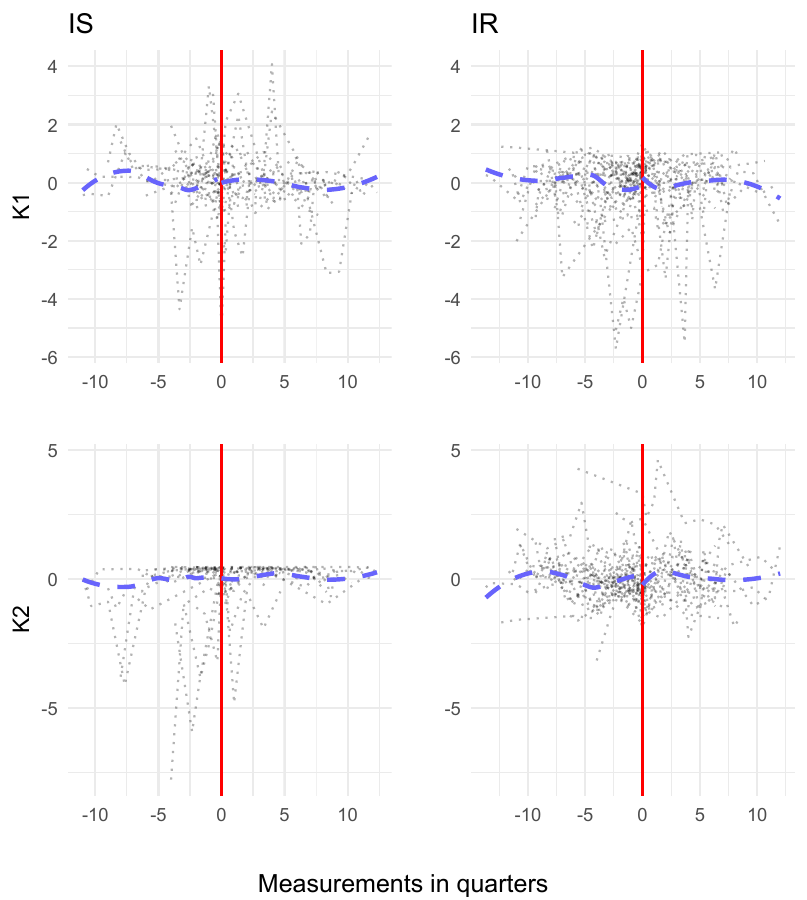}
    \caption{Estimated latent paths for $k=1$ and $k=2$ by group at the order level.}
    \label{fig:hmp-order}
\end{figure}

The latent paths at order level are displayed in figure \ref{fig:hmp-order}. The estimated canonical weights for $k=2$ are somewhat inconclusive at this level. Second component is comprised exclusively of \textit{Deltaproteobacteria}, with negative weight, for the IS group. This bacteria has been found to be more prevalent in prediabetic patients \parencite{delta_chen_2021}. However, it has been suspected to have some anti-inflammatory benefits \parencite{deltaInflam_xia_2021}. On the other hand, the IR group is for once more diversified, including most of the bacteria groups in the composition of the second latent variable (10 out of 12). These difference with the previous levels might be explained by the aggregation level combining changes which where previously dissaggregated.

\subsubsection{Class}
\begin{figure}[!htbp]
    \centering
    \includegraphics[width=0.75\linewidth]{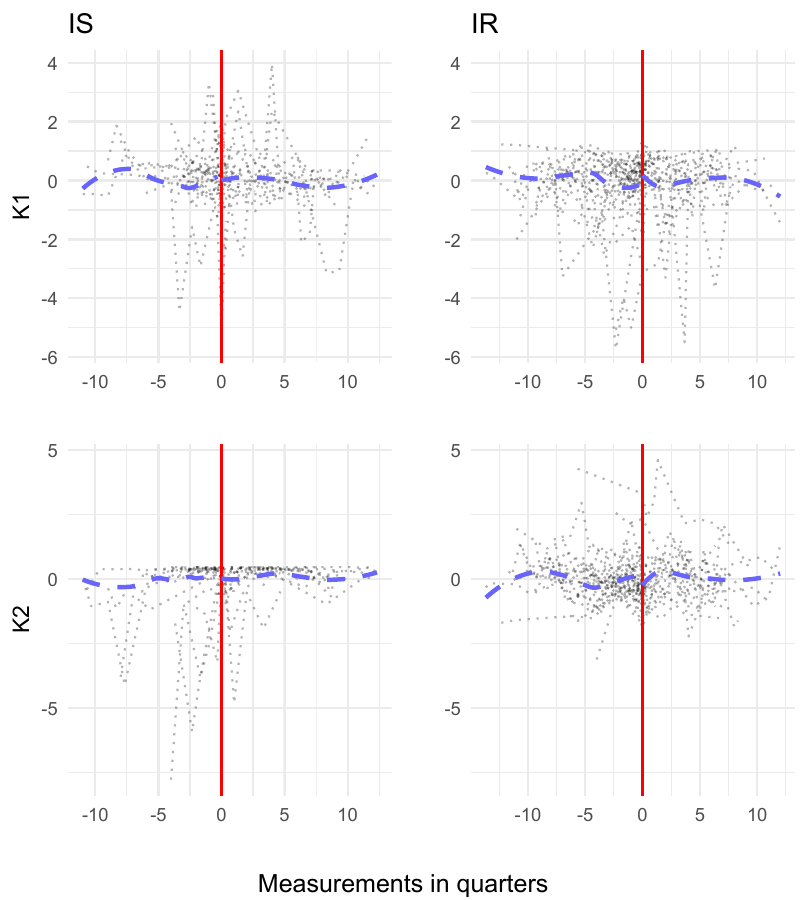}
    \caption{Estimated latent paths for $k=1$ and $k=2$ by group at the class level.}
    \label{fig:hmp-class}
\end{figure}

The estimated latent paths at the phylum level are displayed in figure \ref{fig:hmp-class}. At this level, the analysis is equivalent to the one at order level.  

\subsubsection{Phylum}
\begin{figure}[!htbp]
    \centering
    \includegraphics[width=0.75\linewidth]{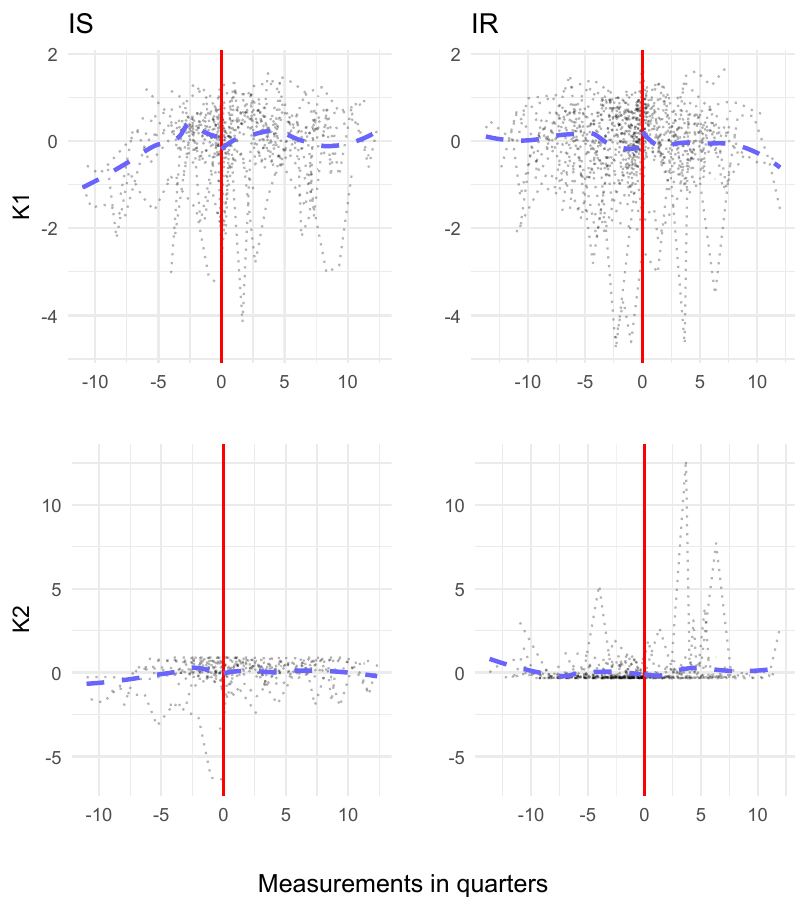}
    \caption{Estimated latent paths for $k=1$ and $k=2$ by group at the phylum level.}
    \label{fig:hmp-phylum}
\end{figure}

The estimated latent paths at the phylum level are displayed in figure \ref{fig:hmp-phylum}. At this level, we observe 6 OTU's. Again, the \textit{Bactiroidetes} and \textit{Firmicutes} groups represent around $90\%$ of gut microbiota in a healthy individual. The \textit{Actinobacteria} and \textit{Proteobacteria} groups are often the ones undergoing larger changes in individuals with disorders affecting the microbiome balance. The estimated canonical weights for $k=2$ gave relevance to the \textit{Actinobateria} OTU which has been consistently linked to IR. For the IS patients we are unable to dive in further as the chosen microbe group was unclassified in the dataset. These results are in line with those presented in the main text. 

\newpage

\end{document}